\documentclass[12pt]{article}
\usepackage{graphicx}
\usepackage{natbib}
\newcommand*{\lm}{{\ell m}}
\newcommand*{\eg}{{\em e.g.}}

\title{Predicting the Polarization of The Microwave Background from the WMAP
Temperature Maps}

\author{A.~H.~Jaffe\\
Astrophysics Group, Blackett Laboratory\\
Imperial College London, SW7 2AZ ENGLAND}

\begin{document}
\maketitle

\begin{abstract}

In this paper, we study how to predict the polarization of the Cosmic
Microwave Background using knowledge of only the temperature (intensity)
and the cross-correlation between temperature and polarization. We
derive a ``Wiener prediction'' method and apply it to the WMAP all-sky
CMB temperature maps and to the MAXIMA field.

\end{abstract}

\section{Introduction}
\label{sec:intro}

Fluctuations in the temperature of the Cosmic Microwave Background (CMB)
have been used to map the surface of last scattering for the last
decade, since the first results from the DMR instrument on the COBE
Satellite \citep{Smoot92}. Now, we have the first results from
WMAP \citep[][references therein and elsewhere in this
volume]{BennettWMAPbasic03} which confirm the CMB anisotropy results
gathered over the intervening decade. But the WMAP data promise more
than just confirmation: they also offer the first high-sensitivity
analysis of the polarization of the CMB \citep{WMAP_TE} [although DASI
can justly be credited with the first detection of cosmological
polarization \citep{DASI_pol02}].

The polarized CMB provides independent information on cosmological
physics; it traces the flow of the plasma at the surface of last
scattering. This flow itself can be decomposed into an irrotational
component, corresponding to the action of gravity by the density
perturbations responsible for the temperature fluctuations, and a
rotational component which can be produced by a background of
gravitational waves (\eg, from inflation). These components result in
patterns of CMB polarization known, respectively, as ``grad'' and
``curl'' \citep{KamKosSte} or, in analogy with electrodynamics, ``E'' and ``B''
\citep{ZalSel97}. The latter (curl or B) thus has the promise of
exploring the physics of the early Universe, but the amplitude of the
signal remains well below current detections. Here, then, we concern
ourselves with grad or E modes.

However, the WMAP team has yet to produce actual maps of the polarized
component of the CMB. Here we investigate a statistical technique to
predict the polarization component based on knowledge of the auto- and
cross-correlations of the components.

\section{Formalism}

In this section, we ask the question: if we know the temperature pattern
on the sky, as well as the cross-correlation between the temperature and
the polarization, what is our best guess at the latter? There are many
ways to tackle this question, mathematically if not philosophically
equivalent. Here, we take a Bayesian approach, and so start with Bayes'
theorem,
\begin{equation}
  \label{eq:bayes}
  P(\theta | D I) = \frac{P(\theta|I) P(D|\theta I)}{P(D|I)}
\end{equation}
where $\theta$ labels the parameters we are trying to determine, $D$ the
data, and $I$ the background information we are bringing to the problem.
$P(a|b)$ is the probability (or probability density) of $a$ given $b$,
so $P(\theta|I)$ is the ``prior'' for $\theta$ and $ P(D|\theta I)$ is
the likelihood --- the probability of getting the specific data $D$
given a fixed set of parameters. The left hand side is the posterior
probability, and the denominator just enforces the normalization
condition: $P(D|I)=\int d\theta P(\theta|I) P(D|\theta I)$.

How do we apply this to the CMB problem? Let us start with a model for
the data, $\Delta_p$ as a sum of signal, $s_p$ and noise, $n_p$,
\begin{equation}
  \label{eq:data}
  \Delta_p = s_p + n_p\; ,
\end{equation}
where $\Delta_p$ gives the value of the data labeled by $p$, which can
include pixel number, frequency of observation, polarization state,
etc. We will take the noise to be described by a zero-mean Gaussian
distribution with
\begin{equation}
  \label{eq:noisevar}
  \langle n_p\rangle = 0;\qquad \langle n_p n_{p'} \rangle=N_{pp'}
\end{equation}
where angle brackets are averages over the distribution of the data, and
$N_{pp'}$ is the noise covariance matrix (which we assume is known, but
see \citet{FerJafMNRAS00,Dore01}). We will also take the signal to be a
zero-mean Gaussian, with analogous properties to those above, and a
signal covariance given by $\langle s_p s_{p'}\rangle=S_{pp'}$. 

Thus the actual likelihood function is given by
\begin{equation}
  \label{eq:Pd}
  P(\Delta | SN) = \frac{1}{\left|2\pi (S+N)\right|^{1/2}} 
  \exp\left[ -\frac12 \Delta^\dagger (S+N)^{-1} \Delta \right]
\end{equation}
where we use matrix notation, $\dagger$ refers to the Hermitian
conjugate, and vertical bars to the determinant. We can then use Bayes'
theorem to ask for the posterior distribution of any quantity related to
the signal, $s_p$.

The most obvious quantity we might estimate is the signal itself,
\begin{eqnarray}
  \label{eq:signal}
  P(s|\Delta SN) &\propto& P(s|SN) P(\Delta|SN) \nonumber\\
  &\propto& \exp[-\frac12 s^\dagger S^{-1} s -\frac12 \Delta^\dagger
  (S+N)^{-1} \Delta]\nonumber\\
  &\propto& \exp\left[ -\frac12 
    \left(s-\langle s|\Delta\rangle\right)^\dagger  C_s^{-1} 
    \left(s-\langle s|\Delta\rangle\right)\right] \;,
\end{eqnarray}
where the second line comes from assigning a Gaussian prior with
variance $S_{pp'}$ to the signal, and the final line comes from
completing the square, and we have defined
\begin{equation}
  \label{eq:sbar}
  \langle s | \Delta \rangle = S (S+N)^{-1} \Delta\qquad\textrm{and}\qquad
  C_s = \langle \delta s\;\delta s^\dagger|\Delta\rangle = S - S(S+N)^{-1}S
\end{equation}
which give the Wiener filter itself, and the variance about the
mean. The Wiener filter has been used in the context of the CMB many
times before \citep[\eg,][]{BunnWiener96}, but primarily as a method of
smoothing or for the separation of various foreground components in the
signal.

But we can ask a somewhat more general question: what is the
distribution of some quantity, $e_p$ for which the prior information is
that $\langle e_p e_{p'}\rangle\equiv E_{pp'}$ and that it is correlated
with the signal, $\langle e_p s_{p'} \rangle = X_{pp'}$? Again, the
result is a Gaussian distribution, this time with mean and variance
given by
\begin{equation}
  \label{eq:ebar}
  \langle e | \Delta\rangle = X (S+N)^{-1} \Delta\qquad\textrm{and}\qquad
  C_e = E - X(S+N)^{-1}X\;
\end{equation}
In fact, all of these results can be compressed into two equations,
using the notation of \citet{BBKS}, already used to some extent above,
\begin{eqnarray}
  \label{eq:genWiener}
  \langle x | d \rangle &=& \langle xd^\dagger\rangle \langle dd^\dagger \rangle^{-1} d
  \;,\qquad\textrm{and}\nonumber\\
  \langle\delta x\;\delta x^\dagger|d\rangle &=& \langle xx^\dagger\rangle -
  \langle xd^\dagger\rangle \langle dd^\dagger \rangle^{-1} \langle
  dx^\dagger\rangle\; ,
\end{eqnarray}
where $d=s+n$ and $x$ is correlated with $s$, so $\langle
xd^\dagger\rangle=\langle xs^\dagger\rangle$, which is the
cross-correlation, $X$, above. If $x=s$, this reduces the original
Wiener filter, Eq.~\ref{eq:sbar}, above.

Information about the variance of the Wiener filter is crucial. The
Wiener mean itself, $\langle x | d \rangle$, is over-smoothed---low in
regimes of low signal-to-noise ratio. A more ``typical'' signal value is
provided by an actual realization from the full posterior distribution:
the mean plus a realization of the correlation matrix.

\section{Applications}

\subsection{Full-sky maps}

The action of the Wiener filter prediction is clearest when we consider
its action on an all-sky map with uniform (or axisymmetric) noise. We
start with a noisy temperature map, which we transform into its
spherical harmonic components, $d_{\ell m}=a_{\ell m} + n_{\ell m}$,
where $a_{\ell m}$ gives the CMB signal and $n_{\ell m}$ the noise. The
Wiener filter and variance are given by
\begin{eqnarray}
  \label{eq:almwiener}
  \langle a_\lm | d_\lm \rangle &=& 
\frac{C_\ell}{B_\ell^2 C_\ell + N_\ell} d_\lm \nonumber\\
\langle E_\lm | d_\lm \rangle &=& 
\frac{C_\ell^{TE}}{B_\ell^2 C_\ell + N_\ell} d_\lm \nonumber\\
\langle\delta a_\lm\delta a_\lm|d_\lm\rangle &=&  C_\ell -
\frac{\left(C_\ell\right)^2}{B_\ell^2 C_\ell + N_\ell}  \nonumber\\
\langle\delta E_\lm\delta E_\lm|d_\lm\rangle &=& C^{EE}_\ell -
\frac{\left(C_\ell^{TE}\right)^2}{B_\ell^2 C_\ell + N_\ell}\nonumber\\
\langle\delta a_\lm\delta E_\lm|d_\lm\rangle &=& C^{TE}_\ell -
\frac{C_\ell^{EE}C_\ell}{B_\ell^2 C_\ell + N_\ell} \;,
\end{eqnarray}
where we have introduced the experimental beam, $B_\ell$, the
temperature power spectrum, $C_\ell=\langle|a_\lm|^2\rangle$, the
E-mode polarization power spectrum, $C^E_\ell=\langle|E_{\ell
  m}|^2\rangle$ and the E-mode polarization/temperature cross power
spectrum, $C_\ell^{TE}=\langle a^*_\lm E_\lm\rangle$. These
power spectra are diagonal in $\ell$, and we impose a similar
constraint on the noise power, which we denote $N_\ell=\langle|n_{\ell
  m}|^2\rangle$. Note that this is an unphysical condition,
corresponding to isotropic noise over the sky; if we restrict ourselves
to high signal-to-noise experiments and large scales. In these
equations, we show the Wiener filter for the temperature itself and for
the polarization, and, by extension of the equations from the previous
section, the covariance between the two.

We note in passing that the formalism also correctly accounts for the
statistics of the polarization ``B-modes''. Since the cross-spectrum
$C_\ell^{TB}=0$, we just have $\langle a^B_\lm|d_\lm\rangle=0$ with
$\langle\delta B_\lm\delta B_\lm|d_\lm\rangle=C_\ell^B$ --- the data
give no new information about them.

We show these formulae for the Wiener filter of the multipole
coefficients themselves, but in fact we will always want to smooth the
resulting map with some beam (although it is not necessary to use the
same as the original experimental beam, $B_\ell$).

We can apply these formulae directly to the temperature data from the
WMAP satellite. Our requirement of isotropic noise should be 
adequate for $\ell<400$ where the noise is negligible. In fact, in the
following we use the Wiener-filtered \emph{temperature} maps provided by
\citet{TegOliHam_WMAP03}. Note that starting from the temperature Wiener
filter enables a slight simplification: 
\begin{equation}
\langle E_\lm|d_\lm\rangle=\frac{C_\ell^{TE}}{C_\ell}\langle a_\lm|d_\lm\rangle
\;.
\end{equation}

\begin{figure}[htbp]
  \centering
  \includegraphics[width=0.75\columnwidth]{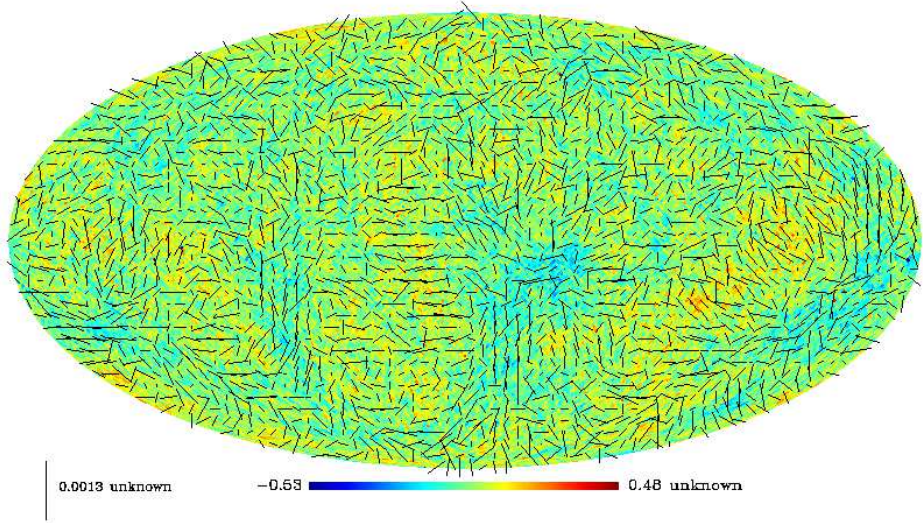}
  \caption{Wiener-filtered map of the CMB temperature and polarization using the WMAP data as prepared by \citet{TegOliHam_WMAP03}.}
  \label{fig:WMAPWiener}
\end{figure}

Starting from these Wiener-filtered data, which are in fact available in
pixel space, we apply these formulae by going back and forth between
pixels and multipole coefficients by performing a spherical-harmonic
transform using routines from the HEALPix package
\citep{healpix}.\footnote{http://www.eso.org/science/healpix/} In
Figure~\ref{fig:WMAPWiener} we show the Wiener-filtered map of the WMAP
data, with the polarization superposed on the temperature.

\begin{figure}[htbp]
  \centering
  \includegraphics[width=0.49\columnwidth]{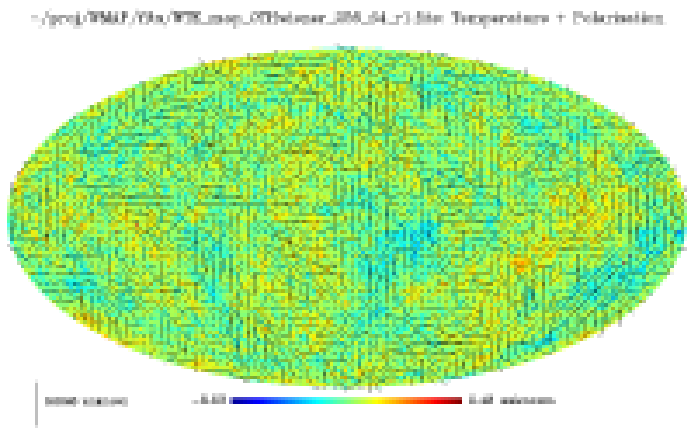}  
  \includegraphics[width=0.49\columnwidth]{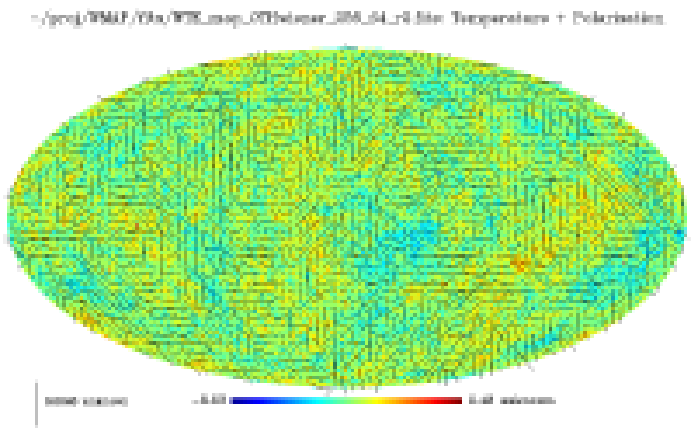}  
  \caption{Two realizations of the map of the CMB temperature and
    polarization using the Wiener-filtered map
    Figure~\ref{fig:WMAPWiener}.} 
  \label{fig:WMAPWienerRlzn}
\end{figure}

\begin{figure}[htbp]
  \centering
  \includegraphics[width=0.8\columnwidth]{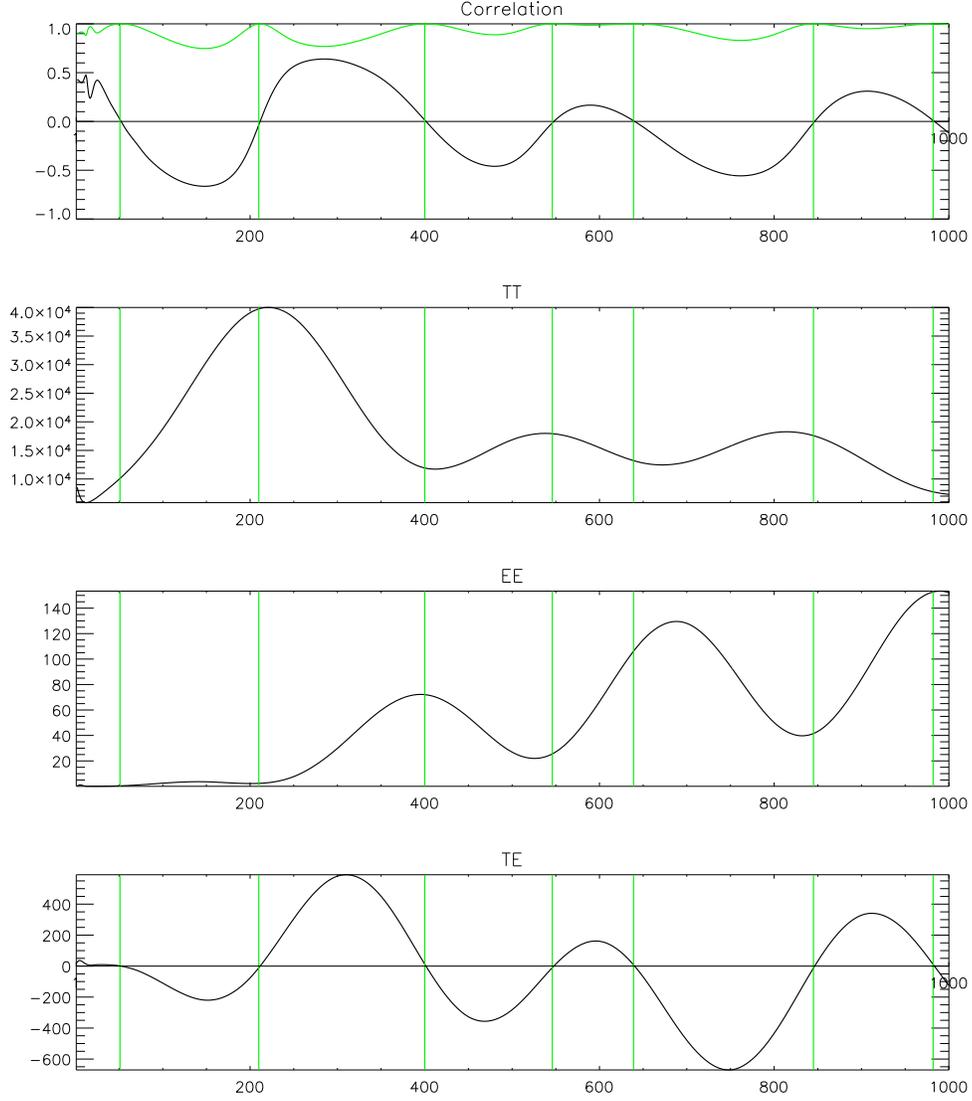}
  \caption{Top Panel: the correlation coefficient, $r$, between 
    temperature and E-mode CMB Fluctuations (bottom curve); the top
    curve shows $1-r^2$. The bottom three panels show the individual
    power spectra, $C_\ell^{XY}$.  The vertical lines give the locations
    where $C_\ell^{TE}=0$. Note that the model we use here has optical
    depth, $\tau=0.17$, responsible for the upturn in the spectra and
    correlation at low $\ell$.}
  \label{fig:corrcoeff}
\end{figure}

In Figure~\ref{fig:WMAPWienerRlzn} we show 4 realizations of the
temperature and polarization drawn from the variances of
Eq.~\ref{eq:almwiener}. It is evident that, in fact, the polarization
field contains a large uncorrelated component. We can understand this by
examining the correlation coefficient,
\begin{equation}
  \label{eq:corr}
  r_\ell^2=\frac{(C_\ell^{TE})^2}{C_\ell^E C_\ell}\;.
\end{equation}
This correlation coefficient gives the fraction of the power at a given
multipole that is correlated between the E-mode polarization and the
temperature.  In the signal-dominated regime (here, low $\ell$) this
gives the correlated fraction of the power in the Wiener filter, as
well.  In addition, the quantity $1-r_\ell^2$ gives the fractional amplitude
of the variance of the Wiener filter. Thus, if $r_\ell\sim1$, a realization
of the Wiener variance is small compared to the Wiener mean, and the
mean is a good indication of the actual signal, but if $r_\ell$ is closer to
$0$, different realizations can look very dissimilar.  We plot the
correlation coefficient in Figure~\ref{fig:corrcoeff}. We see that the
correlated fraction depends strongly on $\ell$ (and indeed drops to zero
where $C_\ell^{TE}=0$) and is everywhere less than about 0.8. In
practice, this means that the realizations do vary considerably, but
this can in principle be ameliorated by careful filtering of the map to
remove those modes where $C_\ell^{TE}\simeq0$.

\subsection{Small areas}

\begin{figure}[htbp]
  \centering
  \includegraphics[width=0.8\columnwidth]{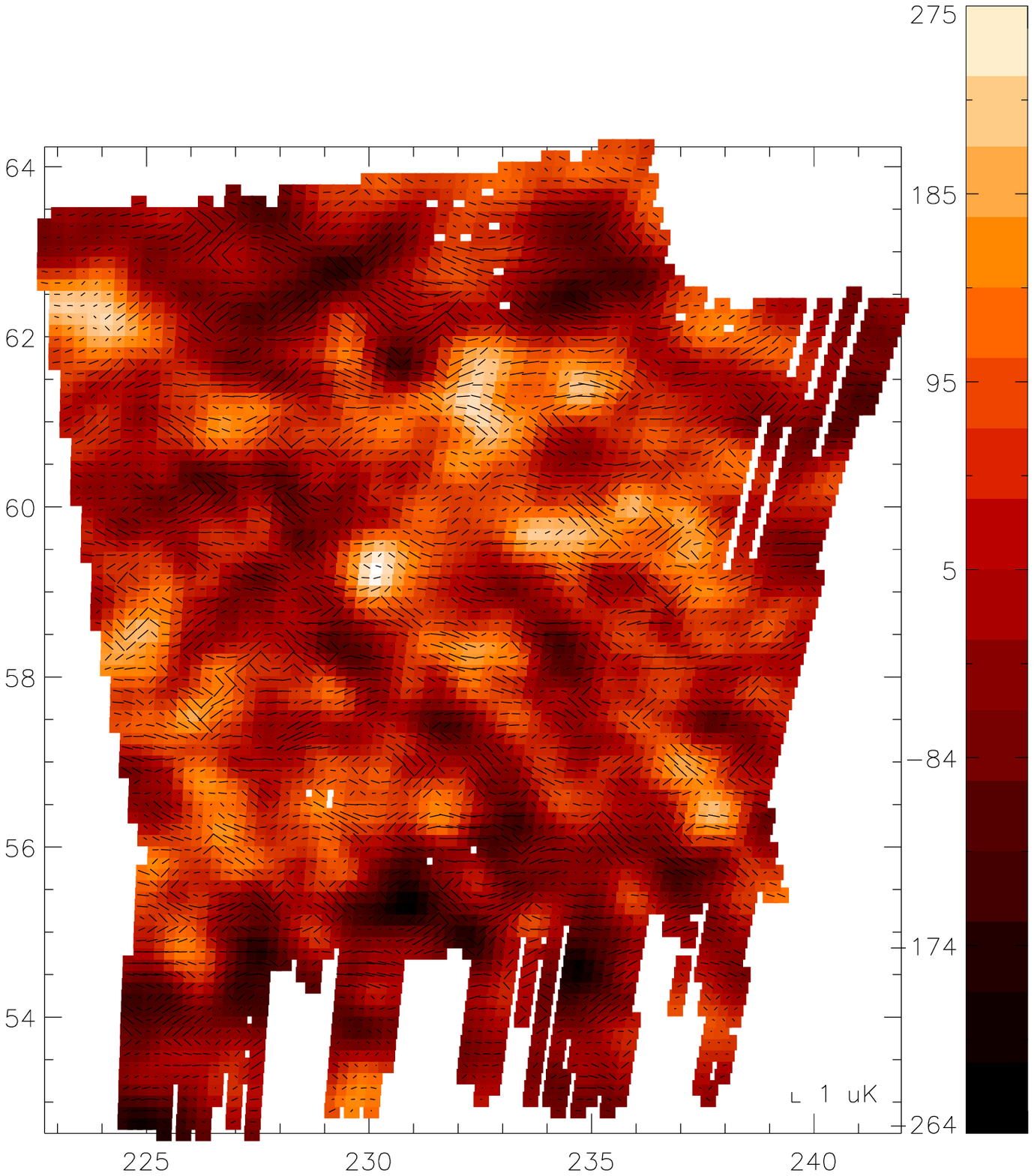}
  \caption{Wiener predicted polarization for the MAXIMA-1 field.}
  \label{fig:MAXIMA}
\end{figure}

We can also use the same technique to predict the polarization for small
areas of sky, where the available data may be more complicated than that
from WMAP. In Figure~\ref{fig:MAXIMA} we show a prediction of the
polarization for the field probed by the MAXIMA experiment
\cite[\eg,][]{Lee01}. Here, we take the full noise and signal covariance
into account, and perform the calculation in pixel space. Instead of
performing the calculation $\ell$-by-$\ell$, we must now perform full
matrix operations. For the isotropic CMB temperature field with power
spectrum $C_\ell$,
\begin{equation}
  \label{eq:SppTemp}
  S_{pp'} = \sum_\ell\frac{2\ell+1}{4\pi} C_\ell B^2_\ell P_\ell(\cos\theta_{pp'})
\end{equation}
where $B_\ell$ gives the spherical harmonic transform of the
experimental beam and $\theta_{pp'}$ is the angle between pixels $p$ and
$p'$. Analogous formulae hold for the covariances of the polarization
components, and slightly more complicated ones for the
cross-correlations amongst and between different polarization and
temperature components.

\section{Discussion}

As mentioned above, the correlation coefficient, $r_\ell$ gives an
indication of the degree to which the the polarization signal is
completely predicted by the temperature signal (and the power spectra
$C_\ell^{XX'}$). The fact that $|r_\ell|<0.8$ is an indication that a
significant fraction of the E-mode polarization signal is not correlated
with the temperature. This leads, finally, to a question about the
physics of the temperature-polarization correlations: the correlated
component of the polarization and temperature is due in large part to
correlations between density and velocity perturbations on the surface
of last scattering \citep{KamKosSte,ZalSel97,jafKamWan00,DodHu02ARAA}. Having only
temperature data as discussed here means that the results \emph{assume}
such a relationship between the two components. However, with the
availability of actual polarization maps, we should be able to use only
minimal theoretical input and yet still recover a two-dimensional
snapshot of the plasma flow on the surface of last scattering. We leave
the details of this to a future work and, we hope, future data.

\subsubsection*{Acknowledgments} 
I would like to acknowledge discussions on this subject with Rob
Crittenden, Lloyd Knox and Dick Bond as well as the COMBAT team at
Berkeley and elsewhere. My involvement with the MAXIMA team has been
invaluable to this work, and in particular has lead directly to
Figure~\ref{fig:MAXIMA}. This work has been supported by PPARC in the
UK.

\def\nat{Nature}
\def\apjl{Astrophys.\ J.\ Lett.}
\def\apj{Astrophys.\ J.}
\def\prd{Phys.\ Rev.\ D}
\def\mnras{Mon.\ Not.\ R.\ Astr.\ Soc.}
\def\aap{Astron.\ Astrophys.}
\def\araa{ARA\&A}
\bibliographystyle{hapalike}
\bibliography{cmb}
\end{document}